\documentclass[prl,twocolumn,showpacs,preprintnumbers,amsmath,amssymb,superscriptaddress]{revtex4}

\usepackage{graphicx}
\usepackage{dcolumn}
\usepackage{bm}

\usepackage{amssymb,amsfonts,amsmath}

\newcommand{\C}{\tilde C(f)}
\newcommand{\R}{\tilde R'(f)}
\newcommand{\Vs}{v_\mathrm{s}}

\newcommand{\Q}{Q_\mathrm{rot}}
\newcommand{\Qs}{Q_\mathrm{s}}
\newcommand{\Qv}{Q_\mathrm{v}}

\newcommand{\uM}{$\mathrm{\mu}\!\!$ M\,\,}
\newcommand{\um}{$\mathrm{\mu}\!\!$ m\,\,}
\newcommand{\PP}{P$_\mathrm{i}$\,}
\newcommand{\Dmu}{\Delta\mu}

\newcommand{\chuo}{Department of Physics, Faculty of Science and Engineering, Chuo University, Tokyo 112-8551, Japan}
\newcommand{\toin}{Faculty of Engineering, Toin University of Yokohama, Kanagawa 225-8502, Japan}
\newcommand{\tokodai}{Graduate School of Bioscience and Biotechnology, Tokyo Institute of Technology, Kanagawa 226-8503, Japan}

\begin{document}

\preprint{}

\title{Nonequilibrium energetics of a single F$_1$-ATPase molecule}

\author{Shoichi Toyabe}
\affiliation{\chuo}
\author{Tetsuaki Okamoto}
\affiliation{\chuo}
\author{Takahiro Watanabe-Nakayama}
\affiliation{\tokodai}
\author{Hiroshi Taketani}
\affiliation{\chuo}
\author{Seishi Kudo}
\affiliation{\toin}
\author{Eiro Muneyuki}
\email{emuneyuk@phys.chuo-u.ac.jp}
\affiliation{\chuo}

\date{\today}

\begin{abstract}
Molecular motors drive mechanical motions utilizing the free energy liberated from chemical reactions such as ATP hydrolysis.
Although it is essential to know the efficiency of this free energy transduction, it has been a challenge due to the system's microscopic scale.
Here, we evaluate the single-molecule energetics of a rotary molecular motor, F$_1$-ATPase, by applying a recently derived nonequilibrium equality together with an electrorotation method.
We show that the sum of the heat flow through the probe's rotational degree of freedom and the work against external load is almost equal to the free energy change per a single ATP hydrolysis under various conditions.
This implies that F$_1$-ATPase works at an efficiency of nearly 100\% in a thermally fluctuating environment.
\end{abstract}

             
                           

\maketitle

\begin{figure}[tbp]
\centerline{\includegraphics[scale=1.0]{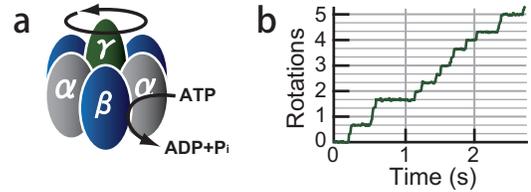}}%
\caption{(Color online) \label{fig:F1}%
{\bf a,} Schematic of F$_1$-ATPase molecule. 
The central $\gamma$ shaft rotates unidirectionally when ATP is hydrolyzed to ADP and phosphate.
$\delta$ and $\varepsilon$ subunits are omitted for simplicity. 
{\bf b,} Typical trajectory at low ATP concentration (0.4 \uM ATP, 0.4 \uM ADP, 1 mM \PP). 
An ATP binding reaction triggers a 120${}^\circ$ step. 
}
\end{figure}

F$_1$-ATPase, a water soluble part of F$_\mathrm{o}$F$_1$-ATP syntheses, is a rotary molecular motor\cite{Boyer1993ah, Abrahams1994,Noji1997,Yasuda1998}.
The central $\gamma$ shaft rotates unidirectionally within a cylinder consisting of three $\alpha$ and three $\beta$ subunits while hydrolyzing ATP to ADP and phosphate (Fig. \ref{fig:F1}a).
A single ATP hydrolysis triggers a 120$^\circ$ rotation \cite{Yasuda1998, Rondelez2005}.
By fixing an F$_1$ motor to a glass surface and attaching a probe filament or particle to the $\gamma$ shaft, we can observe its ATP-driven rotations using conventional optical microscopes at a single molecule level \cite{Noji1997, Yasuda1998, Adachi2007}.
In cells, combined with the membrane embedded proton driven motor F$_\mathrm{o}$, they couple ATP synthesis/hydrolysis and proton flow.
Thus, F$_1$-ATPase plays a central role in biological energy transduction.
Therefore, it is crucial to reveal its energetics, especially the efficiency, to understand the principle of its operation\cite{Muneyuki2007Biophys}.
Although F$_1$-ATPase is known to be highly efficient\cite{Yasuda1998}, well-controlled precise evaluation has not been achieved.
In this study, we evaluated the energetics of the F$_1$-ATPase by measuring thermodynamic quantities of its probe particle using a new nonequilibrium equality \cite{Harada2005, Harada2006, Deutsch2006} together with an electrorotatoin method developed for torque manipulation of microscopic objects\cite{Washizu1991, Berry1995, Watanabe-Nakayama2008}.

In single molecule experiments, what we can access is only the probe attached to molecular motors since motor proteins are quite small with a dimension of around 10 nm.
Accordingly, a methodology to extract energetic information of motors from the probe motion is required.
Although the probe is much larger than the motor protein, it is still sufficiently small that thermal fluctuations have a dominant effect on its behavior.
The probe exchanges energy with an environment through thermal fluctuations as heat \cite{Bustamante2005, Blickle2006}.
%
%
%
Consider a microscopic object moving in a viscous fluid with a frictional coefficient $\Gamma$ and a velocity $v(t)$.
It feels a force of $-\Gamma v(t)+\xi(t)$ from the fluid, where $\xi(t)$ is the thermal fluctuating force.
Then, the heat flow per unit time from the system to the heat bath is naturally defined as $J\equiv\left<\left[\Gamma v(t)-\xi(t)\right]v(t)\right>$, where $\left<\cdot\right>$ is the ensemble average \cite{Sekimoto1997, Sekimoto2010}.
However, since $\xi(t)$ is usually not accessible in experiments, measurement of $J$ in small fluctuating systems has been difficult.
Recently, a nonequilibrium equality that enables us to calculate $J$ from experimentally accessible quantities was derived by Harada and Sasa for Langevin systems in nonequilibrium steady states\cite{Harada2005, Harada2006, Deutsch2006} and experimentally verified\cite{Toyabe2007, Toyabe2008} :
\begin{equation}\label{eq:Harada-Sasa}
J=\Gamma \Vs^2+\Gamma\int^\infty_{-\infty}df\left[\C-2T\R\right],
\end{equation}
where $\Vs$ is the mean velocity.
$\C$ is the fluctuation of the velocity around $\Vs$ calculated as the Fourier transform of the self time correlation function $C(\tau)=\langle[v(t+\tau)-\Vs][v(t)-\Vs]\rangle$.
$\tilde R(f)$ is the Fourier transform of the velocity linear-response function ; $\left<v(t)\right>_\mathrm{N}=\Vs+\int^t_{-\infty}\!\!\! ds\,R(t-s)N(s)+O(N^2)$, where $\left<v(t)\right>_\mathrm{N}$ is the ensemble average of $v(t)$ under a sufficiently small probe torque $N(t)$.
$\tilde R(f)$ indicates the sensitivity of the velocity to a perturbation at $f$.
$\R$ is the real part of $\tilde R(f)$.
$T$ is the temperature.
Boltzmann constant was set to one for the briefness.
Around an equilibrium state, fluctuation response relation (FRR) $\C=2T\R$ is known to hold \cite{Kubo1991}, meaning that a system with large fluctuations is highly sensitive to an external perturbation and vice versa.
On the other hand, FRR is generally violated in nonequilibrium systems\cite{Mizuno2007, Speck2006, Blickle2007}.
Therefore, the magnitude of the integral in (\ref{eq:Harada-Sasa}) indicates how far the system is from equilibrium and vanishes around the equilibrium state.
Thus, the first and second terms of the equality correspond to the dissipation originating in the steady motion and that originating in the nonequilibrium fluctuations, respectively.
It should be noted that heat can dissipate through multiple degrees of freedom, including rotations, swings, radial fluctuations, etc., in a rotational system such as F$_1$-ATPase. 
The equality enables us to track the energy flow in terms of the amount of heat dissipation through a degree of freedom of our interest by measuring the fluctuation and response of that degree of freedom.
In the following, we focus on the rotational degree of freedom of the probe, where $v(t)$ is the rotational velocity and $\Gamma$ the rotational frictional coefficient.

\begin{figure}[tbp]
\centerline{\includegraphics[scale=1.0]{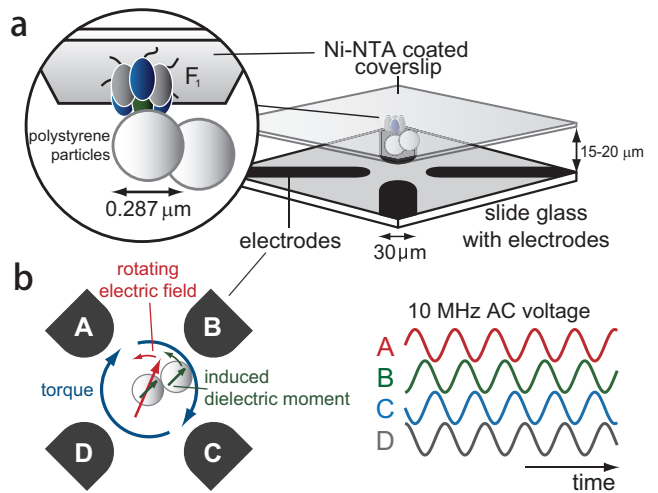}}%
\caption{(Color online) \label{fig:Intro}%
{\bf a,} Experimental setup. 
A dimeric polystyre particle is attached to the $\gamma$ shaft of F$_1$-ATPase fixed on the top glass surface.
Four electrodes are coated on the bottom glass surface.
Distance between electrodes is 50 \um.
{\bf b,} Electrorotation method. 
A rotating electric field at a frequency of 10 MHz was generated at the center of four electrodes by applying sinusoidal voltages with a phase shift of $\pi/2$. 
Dielectric objects in this rotating electric field have a dielectric moment rotating at 10 MHz.
The phase delay between the electric field and the dielectric moment resulted in the torque on the objects.
By periodically modulating the amplitude of the applied voltage, we obtained a sinusoidal torque.
Not to scale.
}
\end{figure}

The experimental setup including the torque calibration was essentially the same as that in \cite{Watanabe-Nakayama2008} (see \cite{EPAPS} for the details).
F$_1$ molecules derived from a thermophilic {\it Bacillus} PS3 with mutations (His$_6$-$\alpha$C193S/W463F, His$_{10}$-$\beta$, $\gamma$S107C/I210C) \cite{Rondelez2005} were fixed on a cover slip functionalized by Ni$^{2+}$-NTA (Fig. \ref{fig:Intro}(a)).
Rotations of a dimeric probe particle with a diameter of 0.287 $\mu$m attached to the $\gamma$ shaft were observed at 24$\pm$1 $^\circ\!$C in a buffer containing  5 mM MOPS-K, indicated amount of ATP, ADP, P$_\mathrm{i}$, and 1 mM excess of MgCl$_2$ over ATP and ADP (pH 6.9) on a phase-contrast microscope (Olympus) equipped with a high-speed camera at 1800 Hz.
Torque necessary for an external load and the response measurement was induced using an electrorotation method\cite{Washizu1991, Berry1995, Watanabe-Nakayama2008} as illustrated in Fig. \ref{fig:Intro}.
The magnitude of torque is constant in the meaning that it does not depend on the particle motion but only on the delay, and is proportional to the square of the amplitude of the applied voltage and the volume of the dielectric objects.
To apply a sinusoidal torque for the response measurement, we modulated the amplitude of the applied 10 MHz AC voltage periodically at various frequencies in the range from 0.3 Hz to 600 Hz.
Temperature rise due to the electric field was at most 0.3${}^\circ$C, which was smaller than the temperature variance of the environment.

\begin{figure}[tbp]
\centerline{\includegraphics[scale=1.0]{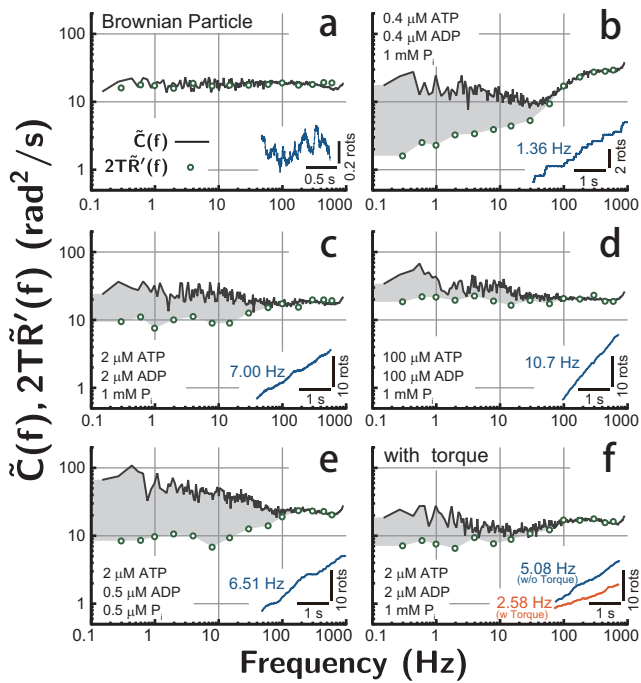}}%
\caption{(Color online) \label{fig:FRR violation}%
Typical examples of fluctuations $\C$ (lines) and responses $2T\R$ (circles) of the rotational velocities.
The shaded area corresponds to the half of the integral in (\ref{eq:Harada-Sasa}).
The rotational trajectory is shown in the inset.
rots : rotations. 
{\bf a,} Rotational Brownian particle in the absence of ATP, ADP, and \PP.
{\bf b--f,} Particles driven by F$_1$-ATPase at indicated concentrations of substrates.
{\bf f,} Under a constant opposing torque of 17.7 pN${}\cdot$nm/rad.
}
\end{figure}

At a low ATP concentration, the probe attached to the $\gamma$ shaft of F$_1$-ATPase rotated unidirectionally with discrete 120${}^\circ$ steps (Fig. \ref{fig:F1}b).
Intervals between steps correspond to the waiting time for ATP binding. 
On the other hand, some particles showed bidirectional rotational Brownian motion.
They were supposed to be attached to collapsed F$_1$ molecules or glass surface through some non-specific bindings.
As a control, we measured $\C$ and $\R$ of such fluctuating particles in a buffer lacking ATP, ADP, and \PP.
Figure \ref{fig:FRR violation}(a) shows that the FRR holds to a good extent, or $\C$ and $2T\R$ coincide well, except in the high frequency region, where system noise, aliasing effects due to the finite sampling rate, and the finite exposure time distort the data.
On the other hand, for continuously rotating particles driven by F$_1$ molecules in a buffer containing ATP, ADP, and \PP, the FRR was violated in the low frequency region (Fig. \ref{fig:FRR violation}(b)).
We suppose that the Lorentzian-like spectrum in the high frequency region reflects Brownian motion in ATP waiting dwells, and the low frequency region reflects stepwise rotation. 
Then, we examined the dependence of $\C$ and $\R$ on ATP, ADP, and \PP concentrations.
As ATP and ADP concentrations were simultaneously increased, average rotational velocity and both $\C$ and $\R$ increased (Fig. \ref{fig:FRR violation}b--d).
At low ATP concentrations, since F$_1$-ATPase spends most of its time trapped in the ATP waiting state, the fluctuation and response are suppressed.
On the other hand at high ATP concentrations, it behaves like a Brownian particle running down a potential gradient, causing the increase in $\C$ and $\R$.
When we held the ATP concentration constant and decreased the ADP and \PP concentrations, the average velocity was nearly the same.
In this case, although $\R$ remained almost constant (Fig. \ref{fig:FRR violation}c,e), $\C$ increased significantly.
This is probably due to the change of fluctuations triggered by the binding and release of ADP and \PP, while an apparent difference in the rotational behavior was not evident.
At last, we applied a constant external torque in the opposing direction of the ATP-driven rotations.
A torque of 17.7 pN$\cdot$nm/rad decreased the mean rotational velocity of a probe from 5.08 Hz to 2.58 Hz (Fig. \ref{fig:FRR violation}f).
Under such an opposing torque, the extent of FRR violation, or the discrepancy between $\C$ and $2T\R$, was less than that in the absence of the  constant opposing torque (Fig. \ref{fig:FRR violation}c, f).


\begin{figure}[bp]
\centerline{\includegraphics[scale=1.0]{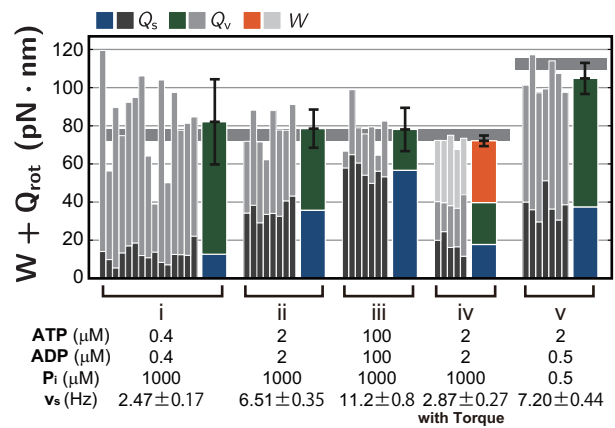}}%
\caption{(Color online) \label{fig:Dissipation}
Amount of heat dissipation and work per 120${}^\circ$ rotation. 
Gray-scale bars correspond to different molecules.
Colored bars correspond to the means. 
The values of the standard free energy difference $\Dmu^\circ$ are different among literatures \cite{ Rosing1972, Guynn1973, Panke1997}.
The horizontal gray bars indicate the range of $\Dmu$ calculated on the basis of values reported in literatures \cite{EPAPS, Rosing1972, Guynn1973, Panke1997} by a method developed by Krab and van Wezel \cite{Krab1992} ; 72.4--78.4 pN$\cdot$nm (i--iv) and 109--115 pN$\cdot$nm (v). 
Error bars indicate the standard deviations. 
}
\end{figure}

From the above results, the amount of heat dissipation through the rotational degree of freedom was estimated using (\ref{eq:Harada-Sasa}).
The integral in (\ref{eq:Harada-Sasa}) corresponds to the twice of the area between $\C$ and $2T\R$ shaded in Fig. \ref{fig:FRR violation} since $\C$ and $\R$ are even functions.
We limited the integration in the range up to 300 Hz to avoid the systematic noise in the high frequency region.
The frictional coefficient $\Gamma$ was calculated from the average of $\C$ around 300 Hz by assuming the FRR in high frequency regions as $\tilde C(f)= 2T/\Gamma$.
The amount of heat dissipation per 120${}^\circ$ rotation was calculated as $Q_\mathrm{rot}=J/3\Vs=\Qs+\Qv$, where $\Qs\equiv \Gamma \Vs^2/3\Vs$ and $\Qv\equiv\Gamma\int df\left[\C-2T\R\right]/3v_s$.
$\Qs$ and $\Qv$ are separately plotted in Fig. \ref{fig:Dissipation}. 
ATP hydrolysis and mechanical motion are known to be tightly coupled in the hydrolytic reactions of F$_1$-ATPase; a single ATP hydrolytic cycle corresponds to a single 120${}^\circ$ step\cite{Yasuda1998, Rondelez2005}.
Thus, the free energy consumption during 120${}^\circ$ rotation is $\Dmu \equiv \mu_\mathrm{ATP}-\mu_\mathrm{ADP}-\mu_\mathrm{P_i} \simeq \Dmu^\circ+T\ln\left(\mathrm{[ATP]}/\mathrm{[ADP][P_i]}\right)$.
In Fig. \ref{fig:Dissipation}(i--iii), the ratio $\mathrm{[ATP]}/\mathrm{[ADP][P_i]}$ and accordingly $\Dmu$ are held to be constant except for a minor difference due to changes in the ionic strength, concentration of free Mg$^{2+}$, etc.
Under these conditions, $\Q$ was almost the same and equivalent to $\Dmu$ shown as a horizontal bar, although $\Qs$ and $\Qv$ varied depending on the ATP concentration.
The increase of ATP concentration resulted in the increase of $\Qs$ and decrease of $\Qv$.
This is consistent with that the $\gamma$ shaft rotates rapidly and smoothly at high ATP concentrations.
When we held the ATP concentration constant at different $\Dmu$, the average velocity and accordingly $\Qs$ were almost the same (Fig. \ref{fig:Dissipation}(ii, v)).
On the other hand, $\Qv$ increased together with $\Dmu$.
As a result, $\Q$ was again comparable with $\Dmu$.
Under a constant external load $N$, F$_1$ performs work $W=N\times 120^\circ$ against the load during $120^\circ$ rotation.
Although $\Qs$ and $\Qv$ decreased simultaneously under an external torque (Fig. \ref{fig:Dissipation}(iv)), the sum $W+\Q$ was almost equal to $\Dmu$.
Thus, we obtained an energy balance relation $W+\Q\simeq\Dmu$ under all the conditions we examined.
In addition, the Stokes efficiency defined as the ratio $\Qs/(\Dmu-W)$ was always less than unity as expected\cite{Wang2001EPL, EPAPS} (Fig. \ref{fig:Dissipation}).
High Stokes efficiency implies that the driving force of a motor is nearly constant.


Since we attached a large probe particle to the $\gamma$ shaft with a relatively soft elastic linker, the time scale of the $\gamma$-shaft motion is much faster than that of the particle.
As the $\gamma$ shaft rotates, it performs a work on the linker, which is stored in the linker as an elastic energy.
This elastic energy then dissipates through the delayed motion of the particle as $\Q$ or used to oppose external load if subjected to external load.
The work on the linker during 120$^\circ$ rotation can not exceed $\Delta\mu$ since it is the work against the conservative elastic force.
Therefore, $W+\Q$ is limited by $\Delta\mu$.
Note that $W+\Q$ is an extractable work by coupling an appropriate system to the probe.
Thus, our result $W+\Q\sim\Delta\mu$ implies that F$_1$ motor works at an efficiency of nearly 100\%.
When the time scale of the probe is faster than that of the $\gamma$ shaft by attaching a sufficiently small probes with a stiff linker, it is expected that the particle  follows the motion of the $\gamma$ shaft without delay.
If this is the case, in addition to $\Delta\mu$, there may be inextractable heat absorption or dissipation through the rigid complex of the probe and shaft.
In our system with a large probe, it is possible that there is such an additional heat flow through the rotation of the $\gamma$ shaft.
However, this heat does not conduct to the particle and is directly exchanged between the shaft and surrounding medium.
Thus, by insulating the probe from the $\gamma$ shaft, we measured only the extractable work of F$_1$-ATPase and found that it works at an efficiency of nearly 100\%.

In summary, we measured the violation of the fluctuation response relation of F$_1$-ATPase at a single molecule level using an electrorotation method.
From the extent of this violation, we calculated the thermodynamic quantities of the probe using a nonequilibrium equality and found that F$_1$-ATPase works at an efficiency of nearly 100\% under a large variety of conditions.
%
The free energy conversion at 100\% efficiency does not contradict the thermodynamic laws, yet it is surprising that such a highly efficient machinery exists and is actually working in cells.
It is also worth noting that F$_1$-ATPase adapts to a large variety of conditoins to retain this efficiency.
Since the rotation of the $\gamma$ shaft is the sole handle to transmit energy between F$_1$ motor and F$_\mathrm{o}$ motor in cells, this high efficiency might have been developed for the efficient coupling of ATP hydrolysis/synthesis and proton transport during a long evolutionary process.
Including this possibility, detailed mechanism and meaning of such highly efficient energy transduction remain a subject for future study.

\begin{acknowledgments}
We appreciate critical discussions with Masaki Sano, Takahiro Harada, Makito Miyazaki, Felix Ritort, Ken Sekimoto, and Kazuhiko Kinosita, Jr.
We thank Shigeru Sugiyama for the technical advice on the electrorotation method.
This work was supported by Japan Science and Technology Agency (JST) and Grant-in-Aid for Scientific Research on Priority Areas, 18074001, 17049015, 19037022, 18031033 (to E.M.), and 21740291 (to S.T.).
\end{acknowledgments}

\bibliography{manuscript}

\end{document}